\documentclass{article}
\usepackage{amscd,amsmath,amssymb,bbm}
\usepackage{amsfonts}
\newcommand{\p}[1]{(\ref{#1})}

\newcommand{\bZ}{{\overline{\strut Z}}{}}
\newcommand{\bXi}{{\overline{\Xi}}{}}

\newcommand{\bD}{{\overline{\strut D}}{}}

\newcommand{\bxi}{{\bar\xi}}
\newcommand{\bpsi}{{\bar\psi}}

\newcommand{\be}{\begin{equation}}
\newcommand{\ee}{\end{equation}}
\newcommand{\bea}{\begin{eqnarray}}
\newcommand{\eea}{\end{eqnarray}}
\newcommand{\ba}{\begin{array}}
\newcommand{\ea}{\end{array}}

\newcommand{\nn}{\nonumber}

\def\={\ =\ }

\def\Nf{$\cal N${=\,}4~}

\begin{document}

\begin{center}
{\Large {\bf${\cal N}=4$ chiral supermultiplet interacting with a magnetic field}}\\[6mm]
{\large {Stefano Bellucci$\,{}^{a}$, Nikolay Kozyrev$\,{}^{b}$,  Sergey Krivonos$\,{}^{b}$, and Anton Sutulin $\,{}^{b}$}}\\
\vspace{1.0cm}
${}^a$ {\it
INFN-Laboratori Nazionali di Frascati,
Via E. Fermi 40, 00044 Frascati, Italy} \vspace{0.2cm}

${}^b$
{\it Bogoliubov  Laboratory of Theoretical Physics,
JINR, 141980 Dubna, Russia}
\vspace{0.2cm}
\end{center}
\begin{abstract}
\noindent We couple \Nf chiral supermultiplet with an auxiliary \Nf fermionic
supermutiplet containing on-shell four physical fermions and four auxiliary bosons. The latter ones play
the role of isospin variables. We choose the very specific coupling which results in a component
action containing only time derivatives of fermionic components presented in the auxiliary supermultiplet,
which therefore may be dualized into auxiliary ones. The resulting component action describes
the interaction of the chiral supermultiplet with a magnetic field constant on the pseudo-sphere
$SU(1,1)/U(1)$. Then we specify the prepotential of our theory to get, in the bosonic sector, the action
for the particle moving over the pseudo-sphere -- Lobachevsky space. We provided also the Hamiltonian
formulation of this system and show that the full symmetry group of our system is $SU(1,1)\times U(1)$.
The currents forming the $su(1,1)$ algebra are modified, as compared to the bosonic case, by the fermionic
and isospin terms, while the additional $u(1)$ current contains only isospin variables.

One of the most important features of our construction is the presence in the Hamiltonian and
supercharges of all currents of the isospin group $SU(2)$. Despite the fact that two of the $su(2)$ currents
$\{ T,{\overline T} \}$ enter the Hamiltonian only through the Casimir operator of the $SU(2)$ group,
they cannot be dropped out, even after fixing the total isospin of the system, because these currents
themselves enter into the supercharges.

We also present the Hamiltonian and supercharges describing the motion of a  particle over
the sphere $S^2$ in the background of constant magnetic field. In this case the additional isospin currents
form the $su(1,1)$ algebra.
\end{abstract}

\setcounter{page}{1}
\setcounter{equation}0
\section{Introduction}
It is a well known fact that the Hamiltonian formalism provides the most natural framework for the description
of quantum mechanics of a isospin-carrying particle in the background of magnetic monopole \cite{mm}.
However, this approach meets a lot of problems when we attempt to add supersymmetry, especially \Nf supersymmetry.
A well known example of such difficulties appears
when we try to find the \Nf supersymmetric extension of the particle moving over the pseudo-sphere
(Lobachevsky space) or the sphere $S^2$ in the field of a Dirac monopole. Indeed, the simplest analysis
shows that the corresponding bosonic Hamiltonian
\be\label{bbos}
{\cal H}_{bos} =
(1- x \bar x)^2 \left( p + \frac{i\bar x}{1- x \bar x}\, J  \right)
\left(  \bar p - \frac{ix}{1- x \bar x}\, J  \right)
\ee
cannot be directly \Nf supersymmetrized within the most natural chiral \Nf supermultiplet
(see e.g. \cite{IS}). As we will show in the present paper, in order to construct the \Nf version of the Hamiltonian
\p{bbos} one should include into the game, together with $J$, all isospin currents spanning $su(2)$ or $su(1,1)$ algebras
in the cases of Lobachevsky space or sphere $S^2$, respectively.

Clearly, the Lagrangian description in terms of superfields is the most suitable for supersymmetric theories.
Therefore, to enjoy all power of the superspace approach one has to use the Lagrangian formalism.
Switching to the Lagrangian description immediately raised the question: which supermultiplet do the isospin degrees of freedom belong to? Indeed, working in the Lagrangian formalism
we have to treat the isospin vector $J$ as a composite one. Thus, the corresponding isospin degrees
of freedom  have to enter the Lagrangian with a kinetic term of the first-order in time derivatives -- only
in this case it will be possible to construct the isospin current as bilinear combinations of isospin
variables after quantization. A first solution of this problem has been proposed in \cite{FIL},
where auxiliary bosonic and fermionic degrees of freedom constitute an auxiliary gauge supermultiplet.
Then this idea has been used  for the construction of \Nf superconformal mechanics with isospin degrees
of freedom \cite{FIL1}. The basic framework for this approach is Harmonic superspace adapted to
one dimension \cite{IL}.

Another approach  \cite{BK1}, which we will exploit  here, uses the ordinary superspace
and ordinary superfields. Its key feature is just a specific coupling
between two \Nf supermultiplets involved in the game, which  gives a kinetic term of the first order in time derivatives for the bosonic isospin variables. In Section 2,
following these arguments, we construct  a coupling
of \Nf chiral supermultiplet and auxiliary \Nf fermionic supermultiplet of this type.
The resulting component action describes the interaction of the chiral supermultiplet with a magnetic
field which is constant on the pseudo-sphere $SU(1,1)/U(1)$. In Section 3 we specify
the prepotential of our theory to get in the bosonic sector the action for the particle moving
over the pseudo-sphere -- Lobachevsky space.
In Section 4 we provide the Hamiltonian formulation of this system explicitly demonstrating
that the constructed system  describes the \Nf supersymmetric isospin carrying particle moving
in the Lobachevsky space and interacting with the constant magnetic field. Our action possesses
the $SU(1,1)$ invariance, similarly to its bosonic analog. In Section 5 we consider
some related problems and extensions. Finally, we conclude  with a summary of results and some possible
further directions of study.

\setcounter{equation}0
\section{Superspace action}
The appropriate way to construct supersymmetric theories is to use a superspace approach in which
the supersymmetry is manifest. In \Nf superspace $\mathbf{R}^{(1|4)} = (t, \theta_i, \bar\theta{}^i)$,
where the spinor derivatives satisfy the standard anticommutation relations
\be\label{alg}
\Big \{ D^i, \bD_j \Big \}=2i\delta^i_j \partial_t, \quad \Big \{ D^i, D^j \Big \}= 0,\quad \Big \{ \bD_i, \bD_j \Big \}=0,
\ee
the chiral supermultiplet is described by a complex scalar superfield $Z$ subjected to the constraints
\be\label{constr1}
D^i Z = 0, \qquad \bD_i \bZ = 0.
\ee
It contains a complex physical boson $\{ z, {\bar z} \}$, a doublet of physical fermions
$\{ \psi_i, \bpsi{}^i\}$ and  auxiliary fields $\{ A, \overline A \}$, which can be defined as
\be\label{comp1a}
z= Z|, \; \bar z= \bZ |, \quad
\psi_i =  \left. \left(\bD_i Z\right)\right|,  \; \bpsi^i =  -\left. \left( D^i \bZ\right)\right|,
\quad A= i \left.\left( \bD_i \bD{}^i Z\right)\right|,\; \overline A= -i \left.\left( D^i D_i \bZ\right)\right|,
\ee
where $|$ in the r.h.s. denotes, as usual, the $\theta=\bar\theta=0$ limit.
The most general $\sigma$-model type superfield action for the chiral supermultiplet is defined through
an arbitrary prepotential $G(Z,\bZ)$ as
\footnote{We define the integration measure to be
 $d^2\theta d^2 \bar\theta = \frac{1}{16}\,D^i D_i  \bD_j \bD{}^j$.}
\be\label{actionchiral}
S_{(\rm Z, \rm \bar Z)}= \int dt d^2\theta d^2 \bar\theta \; G(Z,\bZ).
\ee

One of the possibilities to invent semi-dynamical isospin variables in the Lagrangian of some supersymmetric
mechanics, and therefore to introduce the interaction with magnetic fields, is to
introduce a specific coupling of the basic supermultiplet (i.e. chiral supermultiplet in the case at hands)
with another ``auxiliary'' superfield which contains these isospin variables together with auxiliary fermions.
The most essential feature of such a coupling is that it has to be  a first-order kinetic term for the isospin variables.
The approach, which we will
follow in this Letter, has been elaborated in \cite{BK1,SSA1,KLS1,SSA2}. There, isospin variables and
auxiliary fermions were put in a doublet of fermionic superfields
$\{ \Xi{}^i, \bXi{}^i\}$ subjected to the irreducible conditions \cite{IKL1}
\be\label{constr2}
D^{(i} \Xi^{k)} = 0, \quad \bD^{(i} \Xi^{k)} = 0, \quad
D^{(i} \bXi^{k)} = 0, \quad \bD^{(i} \bXi^{k)} = 0,
\ee
The constraints \p{constr2} leave in the superfields $\{\Xi{}^i, \bXi{}^i\}$ four fermionic and four bosonic components
\be\label{comp1}
\xi^i = \left. \Xi^i \right|,  \; \bar \xi_i =\left.  \bXi_i \right|,  \quad
v=\left.\left(D^i \Xi_i\right)\right|,\; w= \left.\left( \bD_i \Xi^i\right)\right|,\;
\bar w = -\left.\left( D^i \bXi_i\right)\right|,\; \bar v = \left.\left( \bD_i \bXi^i\right)\right|.
\ee
Following \cite{BK1} we introduce the coupling of these chiral superfield  $Z$ and auxiliary one $\Xi^i$ as
\be\label{AI}
S=S_{(\rm Z, \rm \bar Z)}+S_{(\rm Z, \rm \Xi)}=\int dt d^2\theta d^2 \bar\theta \left( G(Z,\bZ)+  F(Z,\bZ) \;\Xi^i \bXi_i\right) \,,
\ee
where $F(Z,\bZ)$ is a real, for the time being arbitrary function.

To ensure that in the action \p{AI} the fermionic fields $\xi^i$ and $\bxi{}_i$ appear only
through the time derivatives and therefore one may replace
the time derivatives of  $\xi^i$ and $\bxi{}_i$ by new fermionic fields $\hat\xi{}^i$ and $\bar{\hat \xi}_i$ as
\be\label{auxFERM}
\hat \xi^i = \dot \xi^i, \qquad
\bar{\hat \xi}_i = \dot{\bar \xi}_i,
\ee
one has to impose some additional constraint on the function $F(Z,\bZ)$. One may check that in the present
case this constraint is incredibly strong
\be\label{analytical}
\frac{\partial^2 F(Z,\bZ)}{\partial Z \partial \bZ} = 0 \quad \Rightarrow \quad F(Z,\bZ) = f(Z) +\overline f(\bZ)\,.
\ee
Moreover, due to the arbitrariness of our prepotential $ G(Z,\bZ)$, without loss of generality, one may assume,
that the function $F(Z,\bZ)$ in \p{analytical} has the form
\be\label{analytical1}
F(Z,\bZ) = Z + \bZ\,,
\ee
and thereby, the superfield action of our model acquires an extremely simple form
\be\label{AModel}
S = \int dt d^2\theta d^2 \bar\theta \left(  G(Z,\bZ) + (Z+\bZ) \;\Xi^i \bXi_i\right)\,.
\ee
After integrating over $\theta's$ in \p{AModel}, replacing the fermionic components \p{auxFERM}
and excluding the auxiliary fields $\{A, \overline A\}$ and $\{ \hat\xi{}^i, \bar{\hat \xi}_i\}$
by their equations of motion
\be\label{eqaux}
A = - \frac{i}{g}\, (w \bar v - g_z \psi^k \psi_k ), \quad
\hat \xi^i = \frac{i}{4(z+\bar z)}\,(v \psi^i - w \bar \psi^i),
\ee
we will get the following on-shell action:
\bea\label{on-shell-action}
S^{(\rm on-shell)} &=& \int dt \bigg \{ g {\dot z}\dot{\bar z}
- \frac{i}{4}\, g \left( \psi_i \dot\bpsi^i -\dot\psi{}_i\bpsi^i\right)
- \frac{i}{4}\, \Big ( g_{\bar z}\dot{\bar z}- g_z {\dot z}\Big ) \psi_i\bpsi^i \nn\\
&+& \frac{i}{8}\,(z+\bar z)(\dot v \bar v - v \dot{\bar v} + \dot w \bar w - w \dot{\bar w})
+ \frac{i}{8}\,(v \bar v -w \bar w)(\dot z - \dot{\bar z})\nn\\
&-&\frac{1}{8(z+{\bar z)}} (v \bar v - w \bar w) \psi_i\bpsi^i
+ \frac{1}{16}\,\Big ( \frac{2}{z+{\bar z}} +\frac{g_z}{g} \Big ) v \bar w \psi^i \psi_i
+ \frac{1}{16}\,\Big ( \frac{2}{z+{\bar z}} +\frac{g_{\bar z}}{g} \Big ) w \bar v \bar \psi_i \bar \psi^i \nn\\
&-& \frac{1}{16g}\,  v \bar v w \bar w
+ \frac{1}{16}\, \Big ( g_{z \bar z} - \frac{g_z g_{\bar z}}{g}\Big ) \psi^i \psi_i \bar \psi_k \bar \psi^k
\bigg \}\,,
\eea
where the metric $g$ is defined as $g = \frac{\partial^2 G(z,\bar z)}{\partial z \partial \bar z}$\,.

The action \p{on-shell-action} is the main result of this Letter. It describes the \Nf supersymmetric
 isospin particle moving in the background of some magnetic field which is specified
by the last term in the second line of the action \p{on-shell-action}. Up to now our consideration was
quite general, and the prepotential $ G(Z,\bZ)$, as well as induced metrics in the configuration
space $g$, were arbitrary. Let us now specify the prepotential to get the cases of special interest
and to clarify the structure of the interaction.
\setcounter{equation}0
\section{Lobachevsky space: Lagrangian}
{}From the explicit form of our superspace action \p{AModel} one could see that the $U(1)$ symmetry,
which seems to be manifest in any theory based on the \Nf chiral supermultiplet, is realized as
\be
\delta Z=i a, \qquad \delta \bZ = - i a.
\ee
Therefore, it makes sense to limit ourselves to considering the prepotential compatible
with this realization. This means that one should choose
\be\label{met1}
G(z,\bar z) = G(z + \bar z) \qquad \Rightarrow \qquad g(z,\bar z)=g(z+\bar z).
\ee
Among the two-dimensional system with the metric \p{met1} there is one special case with
\be\label{Lob1}
g(z, \bar z) = \frac{1}{(z+\bar z)^2}.
\ee
This is just the well known Lobachevsky space. The mechanics on such a configuration space is
commonly used for testing different properties of the theory (see e.g. \cite{ara1,has1}). Let us
analyze our action \p{on-shell-action} for this special case in details.

Making the redefinition of isospin variables as
\be\label{flatcoord}
\tilde v = \frac{\sqrt{z+\bar z}}{2}\,v, \quad \tilde w = \frac{\sqrt{z+\bar z}}{2}\,w, \quad
\bar{\tilde v} = \frac{\sqrt{z+\bar z}}{2}\,\bar v, \quad \bar{\tilde w} = \frac{\sqrt{z+\bar z}}{2}\, \bar w\,,
\ee
and performing the following change of variables:
\bea\label{Poincare}
z&=& \frac{1+x}{1-x}\,, \quad \bar z= \frac{1+\bar x}{1-\bar x}\, \quad \Rightarrow \quad
z+ \bar z = \frac{2(1-x \bar x)}{(1-x)(1- \bar x)}\,, \quad
i\, \frac{\dot z - \dot{\bar z}}{z+\bar z} = - {\cal A} + \dot f\,, \nn\\
\mbox{ where } \quad {\cal A} &=& i\, \frac{\dot x \bar x - x \dot{\bar x}}{1 - x \bar x}\,, \quad
f = i \log \left(\frac{1-\bar x}{1-x}\right)\,, \nn\\
\mbox{ and } \quad
\tilde{\psi}_i &=& \frac{(1-x)(1- \bar x)}{2\sqrt{2}}\,\psi_i\,, \quad
\bar{\tilde \psi}^i = \frac{(1-x)(1- \bar x)}{2\sqrt{2}}\,\bar \psi^i\,,
\eea
one may bring the action \p{on-shell-action} to the form
\bea\label{Poincare-action}
S^{(\rm on-shell)} &=& \int dt \bigg \{\frac{1}{(1 - x \bar x)^2} \dot x \dot{\bar x}
- \frac{i}{2(1 - x \bar x)^2}\,  \Big ( \tilde \psi_i \dot{\bar{\tilde \psi}}^i
-\dot{\tilde \psi}_i \bar{\tilde \psi}^i \Big )
 +\frac{1}{(1 - x \bar x)^2}\, \Big (  {\cal A} - \dot f \Big ) \tilde \psi_i \bar{\tilde \psi}^i \nn\\
&+& \frac{i}{2}\,(\dot{\tilde v} \bar{\tilde v} - \tilde v \dot{\bar{\tilde v}}
+ \dot{\tilde w} \bar{\tilde w} - \tilde w \dot{\bar{\tilde w}})
  -\frac{1}{2}\,(\tilde v \bar{\tilde v} - \tilde w \bar{ \tilde w}) \Big ( {\cal A} - \dot f \Big) \nn\\
&-&  \tilde v \bar{\tilde v} \tilde w \bar{\tilde w}
- \frac{1}{(1 - x \bar x)^2} (\tilde v \bar{\tilde v} - \tilde w \bar{\tilde w}) \tilde \psi_i \bar{\tilde \psi}^i
+\frac{1}{2(1 - x \bar x)^4}\, \tilde \psi^i \tilde \psi_i \bar{\tilde \psi}_k \bar{\tilde \psi}^k
\bigg \}\,.
\eea
Removing the quantity $f$ from the action \p{Poincare-action} by the gauge transformation
\bea\label{1new-Isospin}
&&
\hat v =  e^{-\frac{i}{2} f}  \tilde v\,, \quad \hat w =  e^{\frac{i}{2} f}  \tilde w\,, \quad
\bar{\hat v} =  e^{-\frac{i}{2} f}  \bar{\tilde v}\,, \quad \bar{\hat w} =  e^{\frac{i}{2} f}  \bar{\tilde w}\,,\nn \\
&&
\chi_i = \frac{e^{i f}}{1 - x \bar x}\,  \tilde \psi_i\,, \quad
\bar \chi^i = \frac{e^{-i f}}{1 - x \bar x}\,  \bar{\tilde \psi}^i\,,
\eea
we will finally get the action
\bea\label{Final-action}
S^{(\rm on-shell)}_{final} &=& \int dt \bigg \{\frac{1}{(1 - x \bar x)^2} \dot x \dot{\bar x}
- \frac{i}{2}\,  \left( \chi_i \dot{\bar \chi}^i -\dot{\chi}_i \bar \chi^i\right)
+  {\cal A}  \chi_i\bar \chi^i \nn\\
&+& \frac{i}{2}\,(\dot{\hat v} \bar{\hat v} - \hat v \dot{\bar{\hat v}} + \dot{\hat w} \bar{\hat w} - \hat w \dot{\bar{\hat w}})
-  \frac{1}{2}\,(\hat v \bar{\hat v} - \hat w \bar{\hat w})  {\cal A}  \nn\\
&-&  \hat v \bar{\hat v} \hat w \bar{\hat w}
- (\hat v \bar{\hat v} - \hat w \bar{\hat w}) \chi_i\bar \chi^i
+ (\chi_i \bar \chi^i)^2
\bigg \}\,.
\eea
The action \p{Final-action} describes the motion of the particle in the Lobachevsky space in the
background of a magnetic field ${\cal A}$ with the field strength
\be
F = B(x, \bar x) dx \wedge d \bar x = \frac{2i}{(1- x \bar x)^2} \,dx \wedge d \bar x.
\ee
Clearly, this is just a constant magnetic field on the pseudo-sphere $SU(1,1)/U(1)$ with the metric
$g=\frac{1}{(1 - x \bar x)^2}$.

One should note that the Lobachevsky space, being the coset space $SU(1,1)/U(1)$, possesses
$SU(1,1)$ symmetry. In the Lagrangian form this symmetry is hidden. Moreover, the treatment of the
variables $\{\hat v, \bar{\hat v}, \hat w, \bar{\hat w}\}$ as the isospin ones refers to the
Hamiltonian approach, where they will obey  proper Dirac brackets. In the next Section we will
present the corresponding Hamiltonian and supercharges and analyze the symmetries of our action
\p{Final-action} in full details.
\setcounter{equation}0
\section{Lobachevsky space: Hamiltonian and Symmetries}
In order to find the classical Hamiltonian, we follow the standard procedure for quantizing a system
with bosonic and fermionic degrees of freedom. From the action \p{Final-action} we define the momenta
$\{ p, {\bar p}, \pi{}^i, \bar\pi{}_i, p_v, {\bar p}_v,  p_w, {\bar p}_w\}$ conjugated to
$\{x, {\bar x}, \chi_i,{\bar\chi}{}^i, {\hat v}, \bar{\hat v}, {\hat w}, \bar{\hat w}\}$ respectively,  as
\be\label{momenta}
p=\frac{\dot{\bar x}}{(1 - x \bar x)^2}+\frac{i\bar x}{1 - x \bar x}\left(\chi_i{\bar\chi}{}^i
-\frac{1}{2}(\hat v\, \hat{\bar v} - \hat w \, \hat{\bar w})\right),\quad \pi{}^i =\frac{i}{2} \dot{\bar \chi}^i,\quad
p_v=\frac{i}{2}\, \bar{\hat v},\,\quad p_w=\frac{i}{2}\, \bar{\hat w},
\ee
and introduce Dirac brackets for the canonical variables
\be\label{Dirac}
\left \{x, p \right \} = 1\,, \quad \left \{\bar x, \bar p \right \} = 1\,, \quad
\left \{\chi_i, \bar \chi^k \right \} = i\delta_i^k\,, \quad
\left \{\hat v, \bar{\hat v} \right \} = -i\,, \quad \left \{\hat w, \bar{\hat w} \right \} = -i\,.
\ee
Analyzing the structure of the action \p{Final-action} one may see that the isospin variables enter
the action in rather special combinations. That is why it is convenient to introduce
the following currents $T, \overline T, J$ constructed from the isospin variables
\be\label{TTJ}
T = \hat v \, \hat{\bar w}\,, \quad  \overline T = \hat{\bar v}\, \hat w\,, \quad
J = \frac{1}{2}\,(\hat v\, \hat{\bar v} - \hat w \, \hat{\bar w})\,.
\ee
These currents form the $su(2)$ algebra with respect to the Dirac brackets \p{Dirac}
\be\label{TTJcomm}
\{ T, \overline T \} = 2i J\,, \quad \{ J, T  \} = i T\,, \quad \{ J, \overline T \} = -i \overline T\,.
\ee

Now one may check that the supercharges
\bea\label{QQ}
&&
{\cal Q}_i = (1- x \bar x) p \chi_i  + i \bar x J \chi_i + i \overline T \bar \chi_i
- \frac{i}{2}\, \bar x\, \chi^k \chi_k \bar \chi_i\,,\nn \\
&&
\overline{\cal Q}^i = (1- x \bar x) \bar p \bar \chi^i - i x J \bar \chi^i + i T \chi^i
- \frac{i}{2}\, x\, \bar \chi_k \bar \chi^k \chi^i\,,
\eea
and the Hamiltonian
\be\label{Ham-Lobachevsky}
{\cal H} =
(1- x \bar x)^2 \left( p + \frac{i\bar x}{1- x \bar x}\, \Big(J - \chi_i \bar \chi^i \Big) \right)
\left(  \bar p - \frac{ix}{1- x \bar x}\, \Big(J - \chi_i \bar \chi^i \Big) \right)
+ 2\,J \chi_i \bar \chi^i
- (\chi_i \bar \chi^i)^2 + T\, \overline T\,,
\ee
form the \Nf Poincar\'{e} superalgebra
\be\label{n4Pa}
\left \{{\cal Q}_i, \overline {\cal Q}^k \right \} = i \delta _i^k {\cal H}\,.
\ee
{}From the explicit form of the Hamiltonian one may see that isospin variables enter it only
through the $su(2)$ currents $\{T, \overline T, J\}$ which commute with everything, excluding themselves.
In addition, the  Casimir operator ${\cal C}_{\rm SU(2)}$ of the $SU(2)$ group given by
\be
{\cal C}_{\rm SU(2)} = T\, \overline T + J^2\,,
\ee
commutes with the  Hamiltonian and supercharges. Thus, the relations \p{TTJ}
define the classical isospin matrices. At the same time the fermions enter the
Hamiltonian only through the $u(1)$ current combination $\chi_i \bar \chi^i$
thus providing a description for the fermionic spin degrees of freedom.

Therefore, we conclude that the Hamiltonian \p{Ham-Lobachevsky} indeed describes the motion
of the \Nf supersymmetric isospin carrying particle in the Lobachevsky space in the constant magnetic field.

Now it is time to analyze the symmetries of our Hamiltonian \p{Ham-Lobachevsky}.
In the absence of any interaction the $SU(1,1)$ invariance of the bosonic Hamiltonian is extended
to the supersymmetric case by the following currents:
\be\label{su2without}
R = p - \bar x^2 \bar p + i \bar x \, \chi_i \bar \chi{}^i\,, \;
\overline R = \bar p - x^2 p - i x \, \chi_i \bar \chi{}^i\,, \;
U = i\, (x p - \bar x \bar p)  - \chi_i \bar \chi^i\,.
\ee
These currents span the $su(1,1)$ algebra
\be\label{su11a}
\{ R, \overline R \} = -2i U\,, \quad
\{ U, R \} = i R\,, \quad \{ U, \overline R \} = -i \overline R\,,
\ee
and commute with the supercharges \p{QQ} and the Hamiltonian \p{Ham-Lobachevsky} when $J=T={\overline T} =0$.

It is not too hard to check that the slightly modified currents
\be\label{su2with}
{\cal R} = R - i \bar x J, \quad \overline{\cal R}= \overline R +  i x J, \quad
{\cal U}=U+J
\ee
form the same $su(1,1)$ algebra \p{su11a} and commute now with the full supercharges \p{QQ}
and the Hamiltonian \p{Ham-Lobachevsky}. Thus, our system, describing the \Nf supersymmetric isospin
carrying particle moving in the Lobachevsky space and interacting with the constant magnetic field, possesses
the $SU(1,1)$ invariance, similarly to its bosonic analog. Moreover, if we will introduce
the Casimir operator ${\cal C}_{\rm SU(1,1)}$ of the $SU(1,1)$ group as
\be
{\cal C}_{SU(1,1)} = {\cal R}\, \overline{\cal R} - {\cal U}{}^2\,,
\ee
then one may rewrite our Hamiltonian \p{Ham-Lobachevsky} as  just a sum of two Casimir operators
\be\label{Ham-Casimir}
{\cal H} = {\cal C}_{SU(1,1)} + {\cal C}_{SU(2)}\,.
\ee
It is clear now that the full symmetry of the Hamiltonian \p{Ham-Lobachevsky} is
\be
SU(1,1)\times U(1)  \propto \left\{ {\cal R}, \overline{\cal R},{\cal U} ,J\right\}.
\ee

One should stress that our \Nf system unavoidably  contains all currents of the $su(2)$ algebra
$\{T, \overline T, J\}$. The last term in the Hamiltonian \p{Ham-Lobachevsky} is crucial for
\Nf supersymmetry and cannot be dropped out. That is why the standard procedure of \Nf
supersymmetrization of the bosonic Hamiltonian describing the motion of the particle over the Lobachevsky space
\p{bbos}
fails without introducing additional currents $\{T, \overline T\}$.

It is interesting to note that the total isospin of the system can be fixed as
\be\label{fiso}
{\cal C}_{\rm SU(2)} = T\, \overline T + J^2 =M_0=\mbox{const}.
\ee
With such a fixing the currents $\{T, \overline T\}$ do not show up in the Hamiltonian
\p{Ham-Lobachevsky} (they enter it only through $M_0$) and the system looks like a proper extension
of the "ordinary" bosonic particle interacting with a constant magnetic field. However, the supercharges \p{QQ}
still depend on $\{T, \overline T\}$ and cannot be constructed without such terms.

\setcounter{equation}0
\section{Some steps aside}
Let us consider some relevant questions which are slightly aside our main direction in the paper.
\subsection{Sphere}
The most evident question is what is the situation with a particle moving over $S^2$?
Whether we could introduce the interaction with a constant magnetic field in such a system too,
preserving \Nf supersymmetry? Leaving aside the superspace formulation, one may check that the supercharges
\bea\label{QQSphere}
&&
{\cal Q}_i = (1+ x \bar x) p \chi_i  - i \bar x {\hat J} \chi_i + i \overline{\hat T} \bar \chi_i
+ \frac{i}{2}\, \bar x\, \chi^k \chi_k \bar \chi_i\,,\nn\\
&&
\overline{\cal Q}^i = (1+ x \bar x) \bar p \bar \chi^i + i x {\hat J} \bar \chi^i + i {\hat T} \chi^i
+ \frac{i}{2}\, x\, \bar \chi_k \bar \chi^k \chi^i\,,
\eea
and the Hamiltonian
\be\label{Ham-Sphere}
{\cal H} =
(1+ x \bar x)^2 \left( p - \frac{i\bar x}{1+ x \bar x}\, \Big({\hat J} - \chi_i \bar \chi^i \Big) \right)
\left(  \bar p + \frac{ix}{1+ x \bar x}\, \Big({\hat J} - \chi_i \bar \chi^i \Big) \right)
- 2\,{\hat J} \chi_i \bar \chi^i
+ (\chi_i \bar \chi^i)^2 + {\hat T}\, \overline{\hat T}\,,
\ee
form the \Nf Poincar\'{e} superalgebra \p{n4Pa}, provided the hatted currents $\{ \hat T, \overline{\hat T}, \hat J \}$
span $su(1,1)$ algebra
\be\label{TTJcommhat}
\{ {\hat T}, \overline{\hat T} \} = -2i {\hat J}\,, \quad \{ {\hat J}, {\hat T}  \}
= i {\hat T}\,, \quad \{ {\hat J}, \overline{\hat T} \} = -i \overline{\hat T}\,.
\ee
The Hamiltonian \p{Ham-Sphere} and the supercharges \p{QQSphere} are commuting now with
the generators of the $su(2)$ algebra:
\bea\label{su}
&& {\tilde R} = p + \bar x^2 \bar p + i \bar x \, \chi_i \bar \chi{}^i - i \bar x {\hat J}, \quad
\overline{\tilde R}= \bar p + x^2 p - i x \, \chi_i \bar \chi{}^i +  i x {\hat J}, \nn \\
&&{\tilde U} = i (x p - \bar x \bar p)  - \chi_i \bar \chi^i+{\hat J}.
\eea
Thus the action \p{Ham-Sphere} is invariant under $SU(2)$ transformations  in full analogy with
the \Nf supersymmetric particle in the Lobachevsky space.
\subsection{View from  higher dimensions}
In our Hamiltonian \p{Ham-Lobachevsky} and supercharges \p{QQ} the isospin variables enter through
the $su(2)$ currents $\{ T, \overline{ T},  J \}$ as it should be. All that we need, in order to have a closed
system, is that they commute as in \p{TTJcomm}: the structure of these currents
is irrelevant. This suggests another interpretation of our system. Indeed, let us introduce the additional
complex bosonic field $u$ and the corresponding momenta $p_u$ with the standard brackets
\be
\left\{ u, p_u \right\}=1, \quad \left\{ {\bar u}, {\bar p}_u\right\}=1\;.
\ee
It is clear that now we may realize the currents $\{ T, \overline{ T},  J \}$ as
\be\label{HD}
 T= p_u + {\bar u}{}^2 {\bar p}_u,\quad \overline{ T}= {\bar p}_u + u^2 p_u , \quad
 J = i (u p_u - \bar u {\bar p}_u).
\ee
One may substitute these currents back into the Hamiltonian \p{Ham-Lobachevsky} and supercharges \p{QQ}
to have now a four-dimensional \Nf supersymmetric mechanics. In principle, one may make one step
further and realize the currents $\{  T, \overline{ T},  J \}$ in terms of the complex bosonic
field $u$, the additional real field $\phi$ and their momenta. The resulting system will be five-dimensional.
The detailed analysis of this system will be reported elsewhere.
\setcounter{equation}0
\section{Conclusion}

In the present paper we coupled the \Nf chiral supermultiplet with an auxiliary \Nf fermionic
supermultiplet $\{ \Xi{}^i, \bXi{}^i\}$ containing on-shell four physical fermions and four
auxiliary bosons. These bosons play the role of isospin variables. We choose a very specific
coupling, resulting in a component action which contains only time derivatives of the fermionic
components presented in $\{ \Xi{}^i, \bXi{}^i\}$. This structure of the on-shell action gives the
possibility to dualize these fermions into auxiliary ones, ending up with the  proper action
for matter fields (bosons and fermions) and isospin variables. This procedure was developed in \cite{BK1}.
The conditions selecting such a coupling are very strong and the resulting component action
\p{on-shell-action} describes the interaction of the chiral supermultiplet with a magnetic field
constant on the pseudo-sphere $SU(1,1)/U(1)$. That is why we then specify the prepotential of our theory
to get in the bosonic sector the action for the particle moving over the pseudo-sphere -- Lobachevsky space.
We provided also the Hamiltonian formulation of this system and show that the full symmetry group of
our Hamiltonian is $SU(1,1)\times U(1)$. The currents forming the $su(1,1)$ algebra are modified, as compared
to the bosonic case, by the fermionic and isospin terms, while the additional $u(1)$ current contains only isospin variables.

One of the most important features of our construction is the presence in the Hamiltonian and supercharges of
all currents of the isospin group $SU(2)$  - $\{ T,{\overline T}, J \}$. Despite the fact that the currents
$\{ T,{\overline T} \}$ enter the Hamiltonian only through the Casimir operator of the $SU(2)$ group,
they cannot be dropped out even after fixing of total isospin of the system. The reason for this
behavior is simple: these currents themselves enter into the supercharges. Thus, the \Nf supersymmetry insists
on the presence of these currents in the system. One should stress that these currents appear automatically
from our basic superspace action. Finally, we also present the Hamiltonian and supercharges describing
the motion of a  particle over the sphere $S^2$ in the background of a constant magnetic field.
In this case the additional isospin currents  $\{{\hat T}, \overline{\hat T}, \hat J \}$ form the $su(1,1)$ algebra.

The isospin variables enter our Hamiltonian and supercharges through the $su(2)$ currents
$\{ T, \overline{ T},  J \}$ as it should be. The structure of these currents is irrelevant for
our construction -- it is enough that they span the $su(2)$ algebra. This fact opens a way to build these
currents from additional bosonic variables, thus it gives rise to four- and/or five-dimensional \Nf
supersymmetric mechanics. We leave the analysis of these systems for the future.

The most relevant applications of our result are related with the Quantum Hall effect
on the Lobachevsky space and/or $S^2$ with the constant background magnetic field, which can be thought
of as due to a magnetic monopole seated in the center of a (pseudo) sphere \cite{Haldane,has1}.
Clearly, the system we constructed in this paper provides their \Nf supersymmetric generalizations.
The last step one has to do to analyze the additional effects of \Nf supersymmetry in the Quantum Hall
effect is to construct the full quantum version of
our Hamiltonian and supercharges. We are planning to consider the quantum version of our system in
more details elsewhere.

Finally we note that it is possible and interesting to extend our results to higher dimensional
\Nf mechanics describing, for example, the particle moving over the $CP^n$ manifold in the background of
the magnetic field of some monopole \cite{SSA2}. Indeed, our analysis shows that the \Nf supersymmetrizations we performed
preserve the initial $SU(1,1)$ and/or $SU(2)$ symmetry of the bosonic core. Hopefully, the same will
be true for the system constructed in \cite{SSA3}. We are planning to report our results on the symmetries
of the corresponding \Nf supersymmetric system in a forthcoming paper \cite{Kolya}.

\section*{Acknowledgements}
We are indebted to Armen Nersessian  for valuable discussions.

S.K. and A.S. thank the INFN-Laboratori Nazionali di Frascati, where this work was completed, for
warm hospitality.

This work was partly supported by RFBR grants 11-02-01335,
 11-02-90445-Ukr, as well as by the ERC Advanced
Grant no. 226455, \textit{``Supersymmetry, Quantum Gravity and Gauge Fields''%
} (\textit{SUPERFIELDS}).

\end{document}